\documentclass[sigconf]{acmart}
\AtBeginDocument{%
  \providecommand\BibTeX{{%
    Bib\TeX}}}

\copyrightyear{2025}
\acmYear{2025}
\setcopyright{acmlicensed}\acmConference[SIGMOD-Companion
'25]{Companion of the 2025 International Conference on Management of
Data}{June 22--27, 2025}{Berlin, Germany}
\acmBooktitle{Companion of the 2025 International Conference on
Management of Data (SIGMOD-Companion '25), June 22--27, 2025, Berlin,
Germany}
\acmDOI{10.1145/3722212.3724427}
\acmISBN{979-8-4007-1564-8/2025/06}

\usepackage{algorithmic}
\usepackage{textcomp}
\usepackage{xcolor}
\usepackage{listings}
\usepackage{hyperref}
\usepackage{todonotes}
\usepackage{enumitem}
\usepackage{booktabs} 
\usepackage{multirow} 
\usepackage{subcaption}
\usepackage{diagbox}

\usepackage{cleveref}

\def\BibTeX{{\rm B\kern-.05em{\sc i\kern-.025em b}\kern-.08em
    T\kern-.1667em\lower.7ex\hbox{E}\kern-.125emX}}

\AtBeginDocument{
\crefformat{equation}{#2Eq.~#1#3}
\crefformat{section}{#2Section~#1#3}
\crefformat{subsection}{#2Section~#1#3}
\crefformat{subsubsection}{#2Section~#1#3}
\crefformat{algorithm}{#2Algorithm~#1#3}
\crefformat{figure}{#2Fig.~#1#3}
\crefformat{subfigure}{#2Fig.~#1#3}
\crefformat{table}{#2Table~#1#3}
\crefformat{listing}{#2Listing~#1#3}

\Crefformat{equation}{#2Eq.~#1#3}
\Crefformat{section}{#2Section~#1#3}
\Crefformat{subsection}{#2Section~#1#3}
\Crefformat{subsubsection}{#2Section~#1#3}
\Crefformat{algorithm}{#2Algorithm~#1#3}
\Crefformat{figure}{#2Fig.~#1#3}
\Crefformat{subfigure}{#2Fig.~#1#3}
\Crefformat{table}{#2Table~#1#3}
\Crefformat{listing}{#2Listing~#1#3}
}

\newcommand{\ours}[0]{{\texorpdfstring{\textit{ByteBrain-LogParser}}{ByteBrain-LogParser}}}
\newcommand{\short}[0]{{\texorpdfstring{\textit{ByteBrain}}{ByteBrain}}}

\begin{document}

\lstset{
    language={},
    basicstyle=\ttfamily\small,
    keywordstyle=\color{blue},
    commentstyle=\color{green},
    stringstyle=\color{red},
    breaklines=true,
    columns=fullflexible,
    frame=single,
    captionpos=b,
    showstringspaces=false
}

\title{Adaptive and Efficient Log Parsing as a Cloud Service
}

\author{Zeyan Li}
\affiliation{%
  \institution{ByteDance Inc.}
  \city{Beijing}
  \country{China}
}
\email{lizeyan.42@bytedance.com}

\author{Jie Song}
\affiliation{%
  \institution{ByteDance Inc.}
  \city{Seattle}
  \country{USA}
}
\email{jie.song@bytedance.com}

\author{Tieying Zhang}
\authornote{Corresponding author}
\affiliation{%
  \institution{ByteDance Inc.}
  \city{San Jose}
  \country{USA}
}
\email{tieying.zhang@bytedance.com}

\author{Tao Yang}
\author{Xiongjun Ou}
\affiliation{%
  \institution{ByteDance Inc.}
  \city{Shenzhen}
  \country{China}
}
\email{yangtao.alan@bytedance.com}
\email{ouxiongjun@bytedance.com}

\author{Yingjie Ye}
\affiliation{%
  \institution{ByteDance Inc.}
  \city{Xi'an}
  \country{China}
}
\email{yeyingjie.bluarry@bytedance.com}

\author{Pengfei Duan}
\author{Muchen Lin}
\affiliation{%
  \institution{ByteDance Inc.}
  \city{Chengdu}
  \country{China}
}
\email{duanpengfei.1010@bytedance.com}
\email{linmuchen@bytedance.com}

\author{Jianjun Chen}
\affiliation{%
  \institution{ByteDance Inc.}
  \city{San Jose}
  \country{USA}
}
\email{jianjun.chen@bytedance.com}

\renewcommand{\shortauthors}{Zeyan Li et al.}

\begin{abstract}
Logs are a critical data source for cloud systems, enabling advanced features like monitoring, alerting, and root cause analysis. However, the massive scale and diverse formats of unstructured logs pose challenges for adaptable, efficient, and accurate parsing methods. This paper introduces \ours{}, an innovative log parsing framework designed specifically for cloud environments. \ours{} employs a hierarchical clustering algorithm to allow real-time precision adjustments, coupled with optimizations such as positional similarity distance, deduplication, and hash encoding to enhance performance. Experiments on large-scale datasets show that it processes 229,000 logs per second on average, achieving an 840\% speedup over the fastest baseline while maintaining accuracy comparable to state-of-the-art methods. Real-world evaluations further validate its efficiency and adaptability, demonstrating its potential as a robust cloud-based log parsing solution.  

\end{abstract}


%
  
\maketitle

\section{Introduction}
\label{sec:introduction}
\begin{figure}
\captionsetup{skip=0pt}
    \centering
    \includegraphics[width=\linewidth]{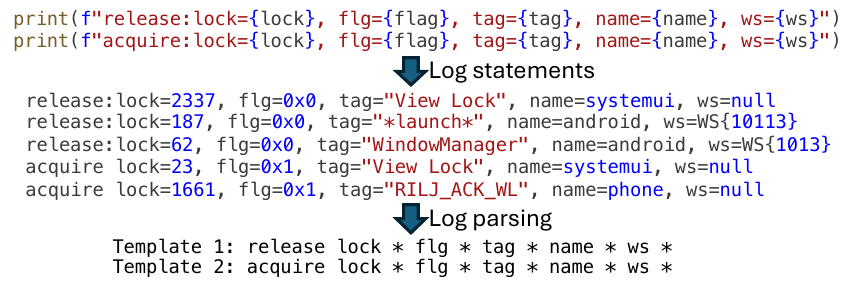}
    \caption{An example of log parsing}
    \label{fig:log-parsing}
    \Description{The figure shows log parsing for "release lock" and "acquire lock" events in an Android system. 
    It extracts two templates, replacing specific values with placeholders (*) for parameters like `lock`, `flags`, `tag`, and `pid`. 
    This demonstrates how logs with similar structures are grouped into templates, even with varying parameter values.}
\end{figure}

System logs provide a rich source of information about the behavior and performance of distributed systems, capturing runtime events, execution flows, and operational states. 
These logs play a pivotal role in automated analysis tasks, including anomaly detection, fault diagnosis, and performance monitoring~\cite{khanGuidelinesAssessingAccuracy2022,liuUniParserUnifiedLog2022,zhaoEmpiricalInvestigationPractical2021,chenPathideaImprovingInformation2022,yaoImprovingStateArtCompression2022}. 
However, the unstructured nature and diverse formats of logs, combined with the vast scale of modern cloud systems, make extracting actionable insights challenging~\cite{chenParserLogParsing2023,zhangSystemLogParsing2023,wangSPINEScalableLog2022}. 
Therefore, most automated log analysis relies on log parsing to automatically extract structured log templates and variables from unstructured log records to address these challenges. \cref{fig:log-parsing} illustrates how parsing transforms code-generated raw log records into structured templates.

Cloud services like AWS \textit{CloudWatch} and Azure \textit{Monitor} provide tenants with foundational log management capabilities, including ingestion, indexing, querying, and basic analytical features.
These platforms process millions of logs per second from diverse application components, aggregating them into massive, heterogeneous streams containing thousands of distinct log templates.
The complexity stems from the diverse nature of modern cloud applications, where each component may generate numerous types of logs for different scenarios and states.
While these platforms provide rule-based grouping, visualization, and alerting, they rely on manual configurations for log parsing, which becomes impractical as template numbers grow.
Recognizing this gap, \textit{Volcano Engine}, the cloud computing services from ByteDance, extends these capabilities by introducing automated, out-of-the-box log parsing, which transforms raw logs into structured formats and enables users to query logs more intuitively.
Based on these parsing results, we further provide multiple advanced analytics capabilities, including log anomaly detection (identifying abnormal changes in template quantities and newly emerged templates), template distribution comparison across different time periods, and automatic matching against a library of known failure scenarios. 
These out-of-the-box features allow users to quickly identify system issues, understand behavioral changes, and diagnose problems.

Delivering log parsing as a cloud service in large scale is fraught with challenges. These include:
\begin{enumerate}[leftmargin=1em]
    \item Adaptability: Applications require varying levels of parsing precision, even for logs within the same stream, necessitating a system that adjusts dynamically to user needs.
    For example, logs generated by \textit{print(register callback for \{\}.format(email))} need be parsed as both \textit{register callback for *} and \textit{register callback for None} during debugging to discover unexpected \textit{null}, while merging these into one template can be better in other scenarios.
    This diversity demands real-time adjustable parsing mechanisms.
    \item Compute Efficiency: 
    The massive scale of cloud logs necessitates efficient algorithms for both offline training and online matching. Delays in log parsing can impede query response time and increase compute resource costs.
    \item Storage Efficiency: Log parsing adds additional storage costs for cloud tenants. 
    To keep the service cost-effective, resource usage must be minimized without compromising performance.
    \item Parsing Accuracy: handling diverse and unpredictable log patterns with high precision, even in the absence of prior knowledge.
\end{enumerate}
Existing techniques inadequately address these challenges. 
Heuristic rule-based methods \cite{heDrainOnlineLog2017,duSpellStreamingParsing2016,jiangAbstractingExecutionLogs2008} struggle to adapt to diverse log patterns.
Frequent pattern mining methods \cite{vaarandiDataClusteringAlgorithm2003,nagappanAbstractingLogLines2010,hamooniLogMineFastPattern2016} are sensitive to parameter tuning and preprocessing, which limits their robustness.
Existing log clustering methods \cite{wangSPINEScalableLog2022,makanjuClusteringEventLogs2009} can fail to generate accurate templates and incur significant overhead.
Deep learning approaches \cite{huoSemParserSemanticParser2023,taoLogStampAutomaticOnline2022} achieve high accuracy but require substantial labeled data and compute resources.
LLM-based methods \cite{jiangLILACLogParsing2024} offer flexibility and adaptability but suffer from high inference costs and latency.
Additionally, most existing works struggle to compute log parsing results at varying precision levels in real-time.

To tackle the challenges of log parsing as a cloud service, we introduce \ours{} (referred to as \short{}), a comprehensive framework tailored for cloud service environments. \short{} is designed to balance adaptability and efficiency, addressing the diverse requirements of cloud tenants. The framework operates in two phases. In the offline training phase, logs are periodically collected and hierarchically clustered into a tree structure, where each node represents a log template. Deeper nodes correspond to more precise templates, enabling granular control over parsing precision. This phase leverages innovative techniques such as positional similarity distance and saturation scoring, which optimize the clustering process to achieve a balance between computational efficiency and parsing precision.

In the online matching phase, \short{} processes incoming logs in real time by matching them against pre-trained templates. The system allows users to adjust parsing precision dynamically, based on their operational needs, by specifying thresholds at query time. This capability enables seamless transition between coarse-grained and fine-grained parsing without requiring the reprocessing of logs. By efficiently managing the interplay between accuracy and resource utilization, \short{} ensures cost-effective operation for high-throughput environments.

Several key techniques enhance \short{}'s efficiency.
Positional similarity distance quantifies log structural similarity through key variable positions, improving clustering accuracy.
Hash encoding enables efficient storage and fast online parsing.
Variable saturation metrics guide hierarchical clustering to optimize template generation and avoid redundant refinements.
These techniques collectively enable \short{} to process diverse, large-scale log streams efficiently.

\begin{figure}
\captionsetup{skip=1pt}
\includegraphics[width=0.95\linewidth]{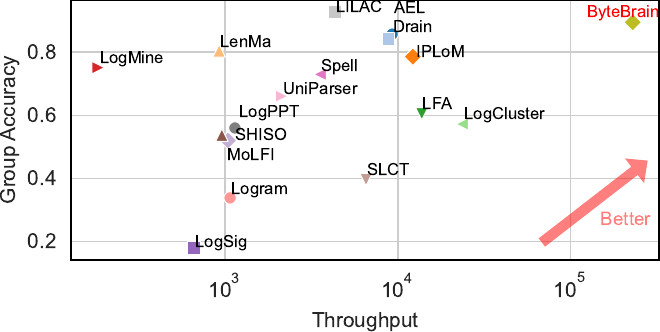}
\caption{Our method meets the goal of high throughput and near-SOTA accuracy}
\label{fig:accuracy-vs-throughput}	
\Description{This figure compares the log parsing accuracy and throughput of various methods on the LogHub-2.0 dataset. 
The x-axis shows throughput (logs per second), and the y-axis shows group accuracy. 
Each method is represented by a point, highlighting its performance in terms of both accuracy and speed. 
\ours{} stands out with high accuracy and significantly higher throughput compared to other methods, demonstrating its superior balance between efficiency and precision.}
\end{figure}

\begin{figure*}[t]
\captionsetup{skip=1pt}
	\includegraphics[width=0.75\linewidth]{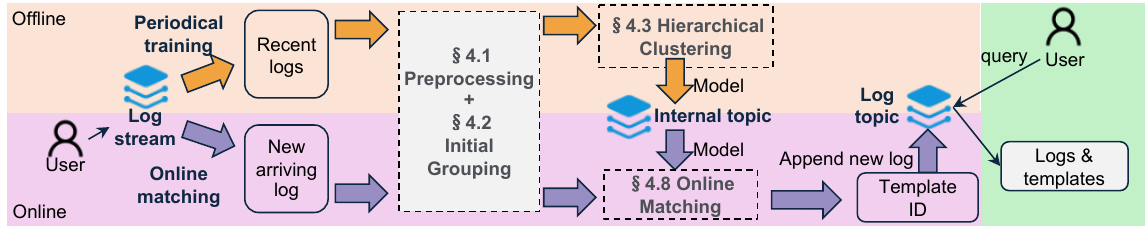}
	\caption{System design of \ours{}}
	\label{fig:system-design}
    \Description{This figure illustrates the system design of \ours{}, which operates in two phases: offline training and online parsing. 
    Logs from each stream are periodically collected and preprocessed, followed by initial grouping and hierarchical clustering to generate a model. 
    During online parsing, new logs are matched against the trained model to identify corresponding templates. 
    The system allows users to query logs and adjust the parsing precision in real time by traversing up the clustering tree to coarser templates when needed.}
\end{figure*}

We evaluate \short{} on the widely-used LogHub and LogHub-2.0 datasets (see \cref{sec:dataset}).  
It achieves an average accuracy of 0.98 and 0.90 on LogHub and LogHub-2.0 respectively, closely matching the SOTA (state-of-the-art) method accuracy of 0.99 and 0.93. 
Its key strength lies in throughput, processing 229,000 logs per second, which is 840.68\% faster than the fastest baseline and outperforms others by 1-3 orders of magnitude.  
As shown in \cref{fig:accuracy-vs-throughput}, this combination of near-SOTA accuracy and excellent efficiency positions \short{} as a highly competitive solution for log parsing as a cloud service.  
Additionaly, our ablation study (\cref{sec:ablation-study}) validates each proposed technique. Industrial evaluations (\cref{sec:industrial-evaluation}) demonstrate its practical advantages in production.
\short{} has been deployed in production on \textit{Volcano Engine}'s \textit{Torch Log Service} (TLS), with real-world evaluations confirming its performance and reliability in cloud computing environments.

In summary, the key contributions of this paper are as follows:
\begin{enumerate}[leftmargin=2em]
    \item An adaptive and efficient log parsing framework for cloud environments: Our system supports real-time, large-scale log parsing and allows users to adaptively adjust parsing precision to meet various operational requirements with low compute and storage overhead. 
    \item An efficient hierarchical clustering-based log parsing algorithm: We introduce positional similarity distance, and variable saturation to enhance log template extraction while minimizing computational overhead and storage costs.
    \item Comprehensive evaluation on large-scale real-world datasets: Extensive experiments demonstrate the near-SOTA accuracy and unprecedented efficiency of our method.
\end{enumerate}

\section{Related Work}

Log parsing remains critical for system management and analysis. Existing techniques are broadly categorized as syntax-based or semantic-based approaches. Each category offers distinct trade-offs in cloud environments where real-time processing, efficiency, and adaptability are paramount. Traditional industrial systems (LogStash~\cite{ElasticLogstashLogstash}, Splunk~\cite{SplunkKeyEnterprise}, CloudWatch~\cite{AmazonCloudWatch}, DataDog~\cite{Datadog}) that rely on user-defined patterns fall outside our focus on automated parsing methods.

Syntax-based log parsers have been widely adopted in log parsing due to their simplicity and compute efficiency.
These parsers primarily rely on predefined rules, heuristics, or patterns to extract structured templates from unstructured log data.
Classic parsers such as \textit{SLCT}~\cite{jiangAbstractingExecutionLogs2008}, \textit{Logram}~\cite{daiLogramEfficientLog2020}, and \textit{Drain}~\cite{heDrainOnlineLog2017} employ various heuristic rules to parse log messages.
For instance, Drain constructs a fixed-depth parse tree for message classification, while Logram leverages n-gram dictionaries for variable token identification.
Another category of syntax-based methods utilizes frequent pattern mining or clustering techniques to identify recurring log structures.
These methods exhibit several limitations: frequent pattern mining-based approaches struggle to identify low-frequency log patterns, while clustering-based methods often suffer from high computational overhead, parameter sensitivity, and suboptimal accuracy.
For example, LogMine's~\cite{hamooniLogMineFastPattern2016} iterative clustering and merging process incurs substantial computational costs, while LogSig~\cite{tangLogSigGeneratingSystem2011} requires precise specification of log category numbers.
LogCluster's~\cite{linLogClusteringBased2016} word-frequency based clustering approach fails to differentiate between semantically distinct messages that share common word distributions.
Notably, although SPINE~\cite{wangSPINEScalableLog2022} employs hierarchical clustering to automatically determine the number of clusters and emphasizes scalability, its log encoding and clustering algorithms impede template generation, limiting its viability as a cloud service.

Semantic-based log parsers have recently emerged, leveraging machine learning and deep learning models to capture deeper dependencies and semantic relationships within log data.
Approaches like \textit{UniParser}~\cite{liuUniParserUnifiedLog2022}, \textit{LogPPT}~\cite{leLogParsingPromptbased2023} and \textit{LogStamp}~\cite{taoLogStampAutomaticOnline2022} represent this category, utilizing custom deep learning models or pretrained language models such as \textit{RoBERTa}~\cite{liuRoBERTaRobustlyOptimized2019} to learn semantic patterns from logs. 
These methods often provide higher parsing accuracy, particularly for complex and unstructured logs, as they do not rely on rigid token-based rules. 
However, semantic-based methods typically demand significant labeled data and compute resources, making them impractical for large-scale or real-time applications. Additionally, their inference costs can result in latency issues, limiting their cost efficiency in cloud environments. 

The rise of Large Language Models (LLMs) offers a promising new direction for log parsing.
LLMs, pre-trained on vast amounts of text data, have demonstrated the ability to understand complex language patterns, making them suitable for parsing log messages without the need for manually crafted rules or extensive labeled data. 
Early work, such as \textit{DivLog}~\cite{xuDivLogLogParsing2024} and \textit{LILAC}~\cite{jiangLILACLogParsing2024}, explores the use of LLMs for log parsing. 
LILAC, in particular, introduces an adaptive parsing cache to mitigate the inefficiency and inconsistency issues commonly associated with LLMs, ensuring both high parsing accuracy and better compute efficiency. 
This represents a significant shift from traditional methods, as it leverages the in-context learning (ICL) capabilities of LLMs to dynamically adapt to different log formats without requiring large-scale retraining. However, these methods still require substantial resources for both pertraining and inference, making them less feasible for large-scale log parsing in cloud environments.

\section{System Design}

\short{} is designed as an adaptive log parsing system tailored for high-throughput cloud environments. 
The system adopts a two-phase approach, including offline training and online matching, to achieve efficiency while maintaining adaptability to diverse log patterns.
\Cref{fig:system-design} illustrates the system architecture.

\textit{Offline Training.} 
A log topic, representing a single log stream, serves as the fundamental unit of our log service, where records are indexed, stored, and made available for analysis. 
Logs within each topic are processed in two phases.
During offline training, logs are collected, preprocessed, initially grouped and hierarchically clustered into a tree structure, where each node represents a log template.
This structure enables \short{} to dynamically adjust parsing precision by traversing the tree based on user-defined thresholds.
The saturation score (see \cref{sec:saturation}), which strictly increases with tree depth, quantifies template precision.  
Each node stores its metadata including template text, saturation score and parent-child relationships in an internal topic. This enables efficient navigation across precision levels while reducing reliance on external databases.  
Training is triggered upon reaching either a volume threshold or a time interval after last execution.
Templates are unavailable for logs before first training completes. However, this limitation is negligible as we configured initial training to finish within 5 minutes, which is inconsequential compared to the typical lifecycle (months to years) of log topics.
For exceptionally large log volumes, random sampling prevents out-of-memory (OOM) issues. The newly trained model is merged with the previous one. Templates with similarity scores above a given threshold are merged; otherwise, they remain separate child nodes.

\textit{Online Matching.} 
The online phase processes incoming logs through preprocessing and initial grouping before matching them against the trained model.
To optimize throughput and latency, the system distributes matching tasks across multiple processing queues, leveraging the independent nature of template matching.
Latency matters because template IDs must be computed along with other traditional text indices before logs can be written to the append-only log topic storage.
For rare log messages that do not appear in the training data, they may fail to match any node in the clustering trees during the online matching phase.
In such cases, we treat the log record itself as a temporary template and insert it into the clustering tree as an individual node.
These unmatched logs are subsequently considered during the next training cycle, allowing the system to adaptively learn new patterns and update the clustering tree accordingly.

\textit{Query.} 
Users can dynamically control template precision during queries by specifying a threshold.
The system efficiently navigates the clustering tree, starting from the retrieved template IDs and traversing upward through ancestor nodes until identifying the coarsest templates that meet the specified threshold.
This approach enables real-time precision adjustment without log reprocessing or redundant template storage, offering adaptive log parsing with minimal computation and storage overhead.

\textit{Parallel.}
The system leverages parallelization across all phases.
During training, preprocessing tasks (tokenization, variable replacement, and hash encoding) and hierarchical clustering operate concurrently.
Moreover, hierarchical clustering can be performed concurrently for each group obtained from initial grouping.
The online phase parallelizes template matching across all logs.
Query processing benefits from parallel execution of both precise template queries and ancestor node traversal.
In production, we optimize resource utilization by limiting parallelization to 1-5 cores, based on topic scale requirements.

\section{Algorithm}
In this section, we introduce our log parsing algorithm.
It starts by transforming unstructured logs into numerical vectors via tokenization and deduplication, followed by initial grouping based on lengths and prefixes to allow parallel processing. 
We then apply hierarchical clustering to each group, with each iteration generating nodes in a tree where deeper nodes represent more precise templates.
We use saturation score to evaluate nodes and determine whether to terminate further clustering.
In \cref{sec:online-matching}, we introduce the online matching algorithm.

\subsection{Preprocessing}
\label{sec:preprocessing}
Preprocessing aims to transform log texts into numerical vectors. This is an important step as it bridges the gap between textual data and mathematical algorithms, which thus enables efficient computation.
Key preprocessing steps include tokenization, common variable replacement, deduplication, and hash encoding.

\subsubsection{Tokenization}
\label{sec:tokenization}

Tokenization refers to the process of dividing each log record into a sequence of tokens. 
By default, we use the following regular expression to segment each log record.

\begin{lstlisting}[caption=Python regular expression for tokenization, label={listing:tokenization}]
(?:://)|(?:(?:[\s\'\";=()\[\]{}?@&<>:\n\t\r,])|(?:[\.](\s+|$))|(?:\\[\"\']))+
\end{lstlisting}

This regular expression comprises four key components: \texttt{://} identifies URL protocol separators; \verb|\s\'\";=()\[\]{}?@&<>:\n\t\r,| captures common delimiters including whitespace, quotes, and punctuation marks; \texttt{[\textbackslash .](\textbackslash s+|\$)} targets sentence-ending periods while preserving those in numerical values; and \verb|\\[\"\']| matches escaped quotation marks frequently used in log data.

We chose regular expressions for tokenization because of their efficiency, simplicity, and customizability.  
While effective tokenization is crucial for successful log parsing, regular expressions can only segment logs based on common delimiters and cannot account for semantic context.  
For example, when processing domain names, whether to use periods as delimiters depends on the specific analysis objectives.  
Nevertheless, regular expressions are fast and allow users to easily define custom tokenization rules for each topic.  
To maintain efficiency, we prohibit the use of high-complexity regex features in user-defined expressions, such as look around, which increases complexity from $O(n)$ to $O(2^n)$ in worst cases. 

\subsubsection{Common Variable Replacement}
\label{sec:regex-replace}
While we focus on automatic log parsing without requiring manual rules, we allow users to optionally specify regex patterns for obvious variables to optimize performance.
These user-defined patterns typically target common, domain-specific variables that appear consistently across logs.
Early replacement of these known variables significantly reduces the complexity for subsequent automatic parsing.
For each topic, we provide default patterns for common variables, including timestamps, IP addresses, MD5 hashes, UUIDs and so on, while users can add domain-specific rules to further enhance efficiency.

\subsubsection{Deduplication}
\label{sec:deduplication}
\begin{figure}[t]
\captionsetup{skip=1pt}
    \centering
    \includegraphics[width=\linewidth]{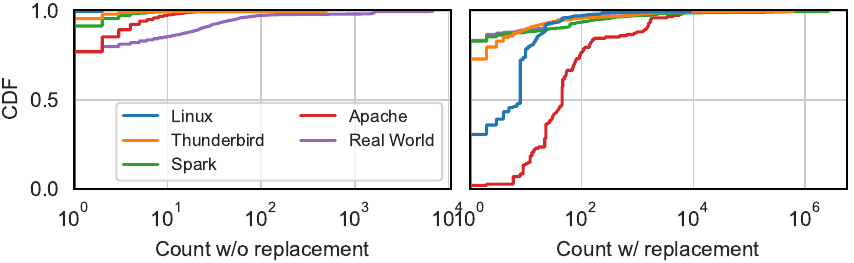}
    
    \caption{High log duplication with increased redundancy after variable replacement}
    \Description{High log duplication with increased redundancy after common variable replacement}
    \label{fig:log_unique_count}
\end{figure}
Log data frequently contains a substantial number of duplicate records, a phenomenon that becomes more pronounced after replacing common variables. 
This redundancy not only increases storage overhead but also introduces inefficiencies in processing and analysis. 
For instance, in \cref{fig:log_unique_count}, we show the distribution of unique log counts across the LogHub 2.0 datasets (see \cref{sec:dataset}). 
The prevalence of repeated log patterns underscores the opportunity for optimization through deduplication.
In this context, deduplication involves identifying and collapsing duplicate log entries while maintaining a count of the occurrences of each unique log statement. 
This approach significantly improves compute efficiency by reducing redundant data.

\subsubsection{Hash Encoding}
\label{sec:hash-encoding}
For compute efficiency, tokens will be encoded into numerical vectors.  
A typical approach is bag-of-words encoding~\cite{wangSPINEScalableLog2022}, which enables Euclidean distance of the encodings are computed and fed into K-means clustering to build log clusters.  
However, this encoding method disregards the order of tokens and cannot directly generate template texts from the clusters.
Alternatively, ordinal encoding, which assigns a unique numerical ID to each distinct token, has a major drawback: it requires storing a mapping between every token and its corresponding numerical ID.  
Given the potentially large number of distinct tokens in logs, this results in substantial storage overhead, significantly increasing the cost of the log parsing service (see \cref{sec:hash-encoding-reduce-space-consumption}).

To address these challenges, we propose to use hash encoding, which leverages a deterministic hash function to map each token to a 64-bit integer.
Using the same hash function for both offline clustering and online matching, we eliminate the need to store token-to-ID mappings.  
Moreover, unlike ordinal encoding, hash encoding supports parallel processing of logs, as the hash function can independently process each token, thereby improving scalability. 

The probability of hash collisions (i.e., two distinct tokens being mapped to the same hash value) is extremely low and can be considered negligible in practice. 
The collision probability can be approximated using the birthday problem formula. 
For a 64-bit hash function, the probability $p$ of at least one collision after hashing $n$ distinct tokens is given by:
\begin{equation}
\begin{aligned}
    p &= 1 - \prod_{k=1}^{n-1}{(1 - \frac{k}{2^{64}})} \approx 1 - \prod_{k=1}^{n-1}{\underbrace{\exp{(- \frac{k}{2^{64}})}}_{\ln{(1-x)}\approx x\text{ when }x\text{ is small}}}\\
    &= 1 - \exp{(\frac{\sum_{k=1}^{n-1})k}{2^{64}}}=1 - \exp{(\frac{n\cdot(n-1)}{2\cdot 2^{64}})}
\end{aligned}
\label{eq:hash-collision}
\end{equation}
For example, with 10 million distinct tokens, the collision probability is only 0.000271\%, which is negligible.
Considering that fields like timestamps and UUIDs are directly replaced with wildcards using regular expressions, the number of distinct tokens processed during the encoding stage is significantly smaller than the number of words in the original log text.

\subsection{Initial grouping}
\label{sec:initial-grouping}
Initial grouping organizes logs into distinct groups based on simple rules to ensure that logs unlikely to belong to the same template are separated early on. 
This allows for more efficient and parallel clustering in later stages.

We apply the following initial grouping strategies:
\begin{enumerate}[leftmargin=2em]
    \item Length: Logs with different token counts are assumed to belong to different templates.
    \item Prefix: Logs are grouped by comparing the first $k$ tokens (configured by users and 0 by default), separating logs with differing prefixes into different groups.
\end{enumerate}

\subsection{Hierarchical Clustering}
\label{sec:hierarchical-clustering}
Hierarchical clustering is a core component of \short{}.
Each initial group serves as the root node of a clustering tree, where the logs are iteratively partitioned into sub-nodes.
At each iteration, the current node is divided into multiple subnodes based on a clustering algorithm tailored for log data.
The clustering process ensures that logs within the same subnode exhibit higher structural similarity (higher saturation as described below) compared to those in the parent node.
The precision of the log templates increases as nodes are partitioned further down the tree.
The tree structure naturally organizes logs, making it easier to identify relationships between templates at different levels of precision.
\Cref{fig:log-cluster-example} below presents clustering tree examples.

The termination of the clustering process for a given node depends on the saturation score, which measures how well the logs in a node have been resolved into either constants or variables.
Nodes with a high saturation score are considered sufficiently refined and do not need further split.
The detailed algorithm for a single clustering process and the calculation of the saturation score are discussed in \cref{sec:single-cluster-process} and \cref{sec:saturation}, respectively.

\subsection{Single Clustering Process}
\label{sec:single-cluster-process}
In the single clustering process, our goal is to iteratively group logs in a manner that ensures the saturation score improves in every cluster. 
This process is inspired by K-Means Clustering and incorporates several modifications to better accommodate the unique characteristics of log data.

The process begins by selecting two logs as the initial cluster centers. 
Following the principles of K-Means++~\cite{arthurKmeansAdvantagesCareful2007}, the first log is chosen randomly, while the second is selected as the log farthest from the first, based on the distance metric described below.
Each of these logs forms the initial core of a cluster, with just one log in each cluster at the start.
Such a strategy prevents similar logs from being incorrectly assigned to different clusters.

To calculate the distance $d(L, C)$ between a log $L$ and a cluster $C$, we propose a positional similarity distance metric instead of the conventional Euclidean distance.  
This choice is driven by the nature of our hash encoding scheme, where token values act as identifiers without meaningful numerical relationships.
Instead, our distance calculation incorporates two key factors: 
\begin{enumerate}[leftmargin=1em]
    \item \textit{Token frequency at each position:} For each token in the log, we evaluate its frequency of occurrence at the corresponding position across all logs in the cluster. A higher frequency indicates that the token is more representative of that position in the cluster. We denote the frequency of the token at position $i$ in log $L$ within cluster $C$ as $f_i(L, C)$.
    \item \textit{Position importance:} Positions with greater variability are more likely to contain variables and are assigned lower importance. We introduce a weight $w_i=\frac{1}{n_i-1}$ for each position $i$, where $n_i$ represents the distinct token count at position $i$ within cluster $C$.
\end{enumerate}
The positional similarity distance is then defined as:
\begin{equation}
    d(L, C)=\frac{\sum_{i=1}^{m}{w_i\cdot f_i(L, C)}}{\sum_{i=1}^{m}{w_i}}
    \label{eq:positional-similarity-distance}
\end{equation}

After computing the distances between each log and the clusters, each log is assigned to the cluster with the smallest distance (i.e., the highest positional similarity). 
This ensures that logs with similar structures are grouped together effectively.

After the initial assignment of logs to clusters, we iteratively refine the clustering results.
In each iteration, as clusters contain new sets of logs, we recalculate the distances between each log and all clusters, then reassign each log to its nearest cluster.
During this process, we monitor the saturation score of each cluster compared to its parent node (the set of all logs).
If a cluster's saturation score shows no improvement, indicating that no additional token positions have been identified as either constants or variables within this cluster, we introduce a new cluster.
The centroid of this new cluster is initialized using the log that exhibits maximum distance from all existing cluster centroids.
This strategic expansion of clusters is naturally bounded by the finite number of token positions in the logs.
Once all possible token positions are classified (reaching a saturation score of 1), further splitting would not yield meaningful improvements.
This approach ensures both adaptability to complex log patterns and compute efficiency by avoiding the creation of unnecessary hierarchical levels.

\subsection{Calculation of Saturation}
\label{sec:saturation}
Saturation is used to evaluate how well the positions in a group of logs have been resolved into constants or variables, and controls the termination of hierarchical clustering.
Unlike existing works~\cite{wangSPINEScalableLog2022}, our saturation calculation considers both confirmed constants and likely variables. It permits faster clustering termination while avoiding unnecessary splits, which improves efficiency and accuracy.

\begin{figure}[t]
    \captionsetup{skip=1pt}
    \centering
    \includegraphics[width=1\linewidth]{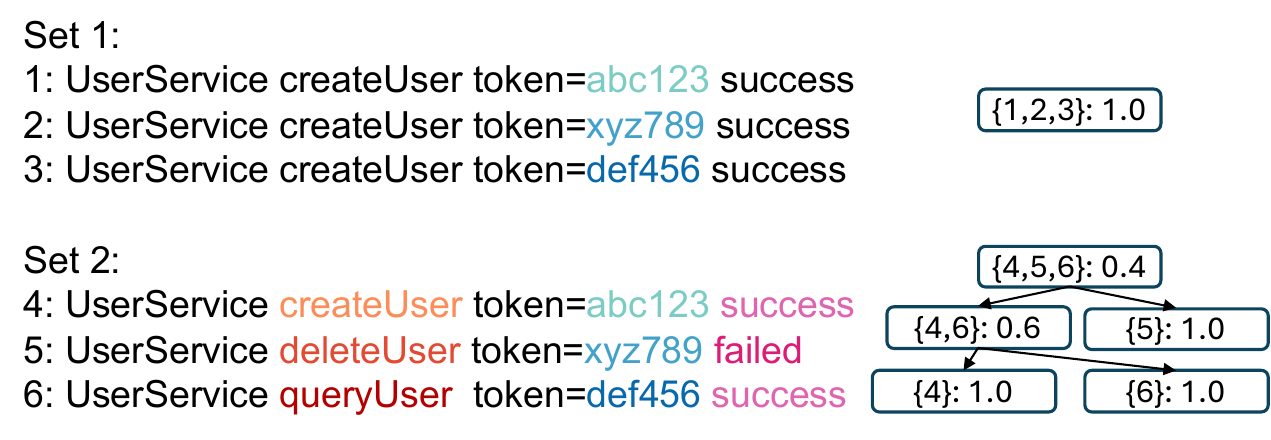}
    \caption{Illustration example of two log sets and the corresponding clustering trees (\textit{{log IDs}: {saturation}} for each node)}
    \Description{The figure shows two sets of UserService operation logs and their corresponding clustering tree. Set 1 consists of three successful createUser operations with different tokens, while Set 2 includes three different operations: a successful createUser, a failed deleteUser, and a successful queryUser. The clustering tree on the right illustrates how these logs are hierarchically grouped, with each node labeled with both the contained log IDs and their saturation values. The tree structure and its branch heights represent the similarity relationships between different log entries, with saturation values ranging from 0.4 to 1.0.}
    \label{fig:log-cluster-example}
\end{figure}

If all logs in a group share the same token at a specific position, that position is definitively considered a constant. 
For example, in \textit{Set 1} of \cref{fig:log-cluster-example}, all positions except the token value are identical across logs, meaning these positions are constants. 
Meanwhile, the high variability of the token position strongly suggests that it is a variable, as further splitting based on token values would not generate meaningful templates.
However, for \textit{Set 2}, though the token position still has different values in every log, there exists value variability across other positions (e.g., the action and status fields). 
This broader variability indicates that the token might not always be a standalone variable but could be structurally correlated with other fields.
In such cases, maintaining separate template for each log preserves important structural patterns.

Therefore, to compute the saturation score, we account for both constant and variable positions in three steps.
\begin{enumerate}[leftmargin=1em]
\item \textit{Proportion of constants }
Let $m$ be the total number of positions in the logs, and $m_c$ the number of positions where all tokens are identical across logs.
The proportion of constants is computed as $f_c = \frac{m_c}{m}$, representing how many positions are fully resolved.

\item \textit{Variability of unresolved positions }
For each unresolved position, let $n$ be the total number of logs and $n_u$ the number of distinct tokens at that position. 
The scale factor for variability at position $i$ is $f_{v}^{(i)} = \frac{\log(n_u) - 1}{\log n}$, which grows with the number of distinct tokens. 
The minimum scale value across all unresolved positions is selected as the overall variability factor $f_v$, ensuring that positions with the highest variability dominate.

\item \textit{Confidence adjustment }
To account for the influence of unresolved positions, we introduce a confidence factor $p_c = \frac{1}{2^{m - m_c - 1}}$, which decreases as fewer positions are resolved. 
This factor ensures that unresolved positions contribute less to the final score when many constants are already identified.
\end{enumerate}

Finally, the saturation score $s(C)$ is computed as:
\begin{equation}
    s(C) = (f_v \cdot p_c + (1 - p_c)) \cdot f_c
    \label{eq:saturation}
\end{equation}
This formula balances the proportion of constants with variability in unresolved positions, giving higher scores to groups that are well-resolved while penalizing groups with high variability.

Based on the saturation definition above, in \cref{fig:log-cluster-example}, for Set 1, the saturation of all three logs is already 1 and thus, we avoid further meaningless splits.
In Set 2, we gradually separate the three logs into different clusters, where the saturation of each new node increases compared to its parent until reaching 1 at the leaf nodes.
Compared to prior methods, our approach provides finer control over the clustering process by dynamically considering both resolved constants and unresolved variability.
This refinement leads to more accurate and meaningful log templates and enhances efficiency by avoiding meaningless splits.

\subsection{Balanced grouping}
\label{sec:balanced-group}
When calculating the distance between a log and multiple clusters, it is common for the log to have the same distance to multiple clusters. 
In such cases, to ensure even cluster distribution, we randomly assign the log to one of these clusters with equal probability, rather than deterministically assigning it to the first cluster. 

Balancing the distribution of logs across clusters helps minimize the depth of the resulting clustering tree.
When users specify a saturation threshold at query time, the system identifies the coarsest template that satisfies the threshold by traversing the ancestor nodes of the most precise template (pre-computed during online matching). 
A shallower tree means fewer nodes need to be traversed, which improves query efficiency and reduces latency.
It could also reduces the total number of iterations required by the clustering algorithm, thereby improving compute efficiency at training time.

\subsection{Early Stop}
\label{sec:early-stop}
In certain cases, the algorithm can immediately determine that each distinct log should form a separate cluster without proceeding with the full clustering process. 
Ending early in these situations could reduce the computational overhead. 
Early stop applies in the following scenarios:
\begin{enumerate}[leftmargin=1em]
    \item \textit{Few logs:} If the number of logs is less than or equal to 2, each log naturally forms a separate cluster.
    \item \textit{Single unresolved position:} If only one position remains unresolved (i.e., it cannot be classified as a constant or variable), further clustering is unnecessary since splitting based on a single position will not increase saturation.
    \item \textit{Completely distinct unresolved positions:} If unresolved positions contain entirely different tokens in each log, these logs are inherently dissimilar and should belong to separate clusters.
\end{enumerate}

\subsection{Online Matching}
\label{sec:online-matching}

In the online matching process, logs are directly matched to template texts instead of traversing the clustering tree by recalculating positional similarity distances. 
This approach significantly reduces storage requirements compared to recalculating distances at each tree node, which would require storing detailed token-level information (e.g., token frequencies or variability metrics) for every node. 
By storing only the final template texts, our method avoids this overhead, making it more efficient for cloud-based log parsing.

When a new log arrives, it is sequentially matched against all templates in descending order of saturation score, stopping as soon as a match is found. 
The matching process is position-based: the token in the log must either match the token in the template exactly or match a wildcard, which indicates a variable.
Although this method does not guarantee that a log will map to the exact node it would have been assigned to during clustering, it achieves high accuracy. 
This is because the templates, generated through clustering, already capture the key structural patterns of logs, and the saturation score prioritizes templates that are both precise and general.
It is also demonstrated by our experiment results in \cref{sec:ablation-study}.
In summary, this approach ensures accurate and efficient matching without recalculating distances or traversing the tree.

\section{Experiment}
\label{sec:experiment}
In this section, we comprehensively evaluate the effectiveness and efficiency of \short{} on widely-used public datasets.


\subsection{Experiment Setup}
\subsubsection{Dataset}
\label{sec:dataset}

\begin{table}[t]
\captionsetup{skip=1pt}
\centering
\setlength{\tabcolsep}{2pt} 
\footnotesize 
\caption{Loghub and Loghub-2.0 dataset statistics}
\label{tbl:datasets}
\begin{tabular}{l|rrr|rrr}
\toprule
& \multicolumn{3}{c|}{Loghub} & \multicolumn{3}{c}{Loghub-2.0} \\
Dataset Name & \#Logs & Size & \#Templates & \#Logs & Size & \#Templates \\
\midrule
HealthApp & 2000 & 183.06 KB & 75 & 212394 & 19.53 MB & 156 \\
OpenStack & 2000 & 581.17 KB & 43 & 207632 & 58.56 MB & 48 \\
OpenSSH & 2000 & 219.94 KB & 27 & 638947 & 67.27 MB & 38 \\
Proxifier & 2000 & 231.41 KB & 8 & 21320 & 2.40 MB & 11 \\
HPC & 2000 & 147.63 KB & 46 & 429988 & 31.10 MB & 74 \\
Zookeeper & 2000 & 273.33 KB & 50 & 74273 & 9.85 MB & 89 \\
Mac & 2000 & 311.93 KB & 341 & 100314 & 14.72 MB & 626 \\
Hadoop & 2000 & 375.93 KB & 114 & 179993 & 30.41 MB & 236 \\
Linux & 2000 & 211.41 KB & 118 & 23921 & 2.04 MB & 338 \\
Android & 2000 & 272.54 KB & 166 & - & - & - \\
HDFS & 2000 & 281.10 KB & 14 & 11167740 & 1.46 GB & 46 \\
BGL & 2000 & 309.72 KB & 120 & 4631261 & 686.12 MB & 320 \\
Windows & 2000 & 278.74 KB & 50 & - & - & - \\
Apache & 2000 & 167.23 KB & 6 & 51978 & 4.75 MB & 29 \\
Thunderbird & 2000 & 317.57 KB & 149 & 16601745 & 2.34 GB & 1241 \\
Spark & 2000 & 191.67 KB & 36 & 16075117 & 1.52 GB & 236 \\
\bottomrule
\end{tabular}
\end{table}

We evaluate our method on two widely-used public datasets: LogHub~\cite{zhuLoghubLargeCollection2023} and LogHub-2.0~\cite{jiangLargescaleBenchmarkLog2023}. 
The original LogHub dataset~\cite{zhuLoghubLargeCollection2023}, consisting of 16 diverse datasets from various sources such as distributed systems, operating systems, and software applications, has been widely adopted in numerous log parsing studies~\cite{daiLogramEfficientLog2020,wangSPINEScalableLog2022,chuPrefixGraphVersatileLog2021,liuUniParserUnifiedLog2022,yuBrainLogParsing2023,zhangSystemLogParsing2023}. 
However, its relatively small size (2,000 labeled logs per dataset) limits its validity. 
To address this, LogHub-2.0 extends LogHub with larger-scale labeled logs, some exceeding 50 million messages, and has gained traction as a benchmark for scalable log parsing~\cite{jiangLILACLogParsing2024}. 
The combination of these datasets allows us to evaluate our method comprehensively, assessing both parsing accuracy on diverse sources and efficiency on large-scale logs.

\subsubsection{Baselines}
We compare against a comprehensive set of baseline methods, covering diverse log parsing techniques.  
Syntax-based baselines include clustering-based approaches, such as 
IPLoM~\cite{makanjuClusteringEventLogs2009}, LogCluster~\cite{linLogClusteringBased2016}, and LenMa~\cite{shimaLengthMattersClustering2016}; frequent pattern mining approaches, such as SLCT~\cite{vaarandiDataClusteringAlgorithm2003}, LFA \cite{nagappanAbstractingLogLines2010}, LogMine~\cite{hamooniLogMineFastPattern2016}, and SHISO~\cite{mizutaniIncrementalMiningSystem2013}; heuristic rule-based approaches, such as AEL~\cite{jiangAbstractingExecutionLogs2008}, Drain~\cite{heDrainOnlineLog2017} and Spell~\cite{duSpellStreamingParsing2016}; search-based approach, including Logsig~\cite{tangLogSigGeneratingSystem2011} and MoLFI~\cite{messaoudiSearchBasedApproachAccurate2018}.  
Deep learning-based baselines include UniParser~\cite{liuUniParserUnifiedLog2022}, Logram~\cite{daiLogramEfficientLog2020} and LogPPT~\cite{leLogParsingPromptbased2023}.
LLM-based methods are represented by LILAC~\cite{chenParserLogParsing2023}.
These baselines are implemented with either the open-source Logparser toolkit~\cite{zhuToolsBenchmarksAutomated2019} or their official open-source code.

\subsubsection{Evaluation Metrics}
We adopt the following standard metrics to evaluate the performance of our method, consistent with prior work in log parsing~\cite{heDrainOnlineLog2017,wangSPINEScalableLog2022,yuBrainLogParsing2023,chenParserLogParsing2023,liuUniParserUnifiedLog2022}.
\begin{itemize}[leftmargin=1em]
    \item Grouping Accuracy (GA): The ratio of correctly grouped logs to total logs. A log is correctly grouped only when placed with all other logs sharing its ground-truth template. This strict metric prevents accuracy inflation from simple, frequent patterns.
    
    \item Throughput (logs/sec): Logs processed per second, calculated as the total log count divided by the combined time for model training and log matching.
\end{itemize}

\subsection{Effectiveness Comparison}
\label{sec:effectiveness-comparison}

\begin{table*}[htbp]
\captionsetup{skip=3pt}
\centering
\setlength{\tabcolsep}{1pt} 
\footnotesize 
\caption{Group Accuracy Comparison on LogHub. The highest group accuracy for each dataset is highlighted in bold, and the second highest is underlined.}
\label{tbl:group-accuracy-2k}
\begin{tabular}{lccccccccccccccccc}
\toprule
Method & Android & Apache & BGL & HDFS & HPC & Hadoop & HealthApp & Linux & Mac & OpenSSH & OpenStack & Proxifier & Spark & Thunderbird & Windows & Zookeeper & Average \\
\midrule
AEL & 0.68 & \textbf{1.00} & 0.76 & \textbf{1.00} & 0.90 & 0.54 & 0.57 & 0.67 & 0.76 & 0.54 & 0.76 & 0.52 & 0.91 & 0.94 & 0.69 & 0.92 & 0.76±0.17  \\
Drain & 0.91 & \textbf{1.00} & 0.96 & \textbf{1.00} & 0.89 & 0.95 & 0.78 & 0.69 & 0.79 & 0.79 & 0.73 & 0.53 & 0.92 & 0.96 & \textbf{1.00} & 0.97 & 0.87±0.14  \\
IPLoM & 0.71 & \textbf{1.00} & 0.94 & \textbf{1.00} & 0.82 & 0.95 & 0.82 & 0.67 & 0.67 & 0.80 & 0.87 & 0.52 & 0.92 & 0.66 & 0.57 & 0.96 & 0.80±0.15  \\
LenMa & 0.88 & \textbf{1.00} & 0.69 & \textbf{1.00} & 0.83 & 0.89 & 0.17 & 0.70 & 0.70 & 0.93 & 0.74 & 0.51 & 0.88 & 0.94 & 0.57 & 0.84 & 0.77±0.21  \\
LFA & 0.62 & \textbf{1.00} & 0.85 & 0.82 & 0.55 & 0.89 & 0.90 & 0.28 & 0.60 & 0.50 & 0.20 & 0.03 & \underline{0.99} & 0.65 & 0.59 & 0.84 & 0.64±0.29  \\
LogCluster & 0.80 & 0.71 & 0.83 & 0.79 & 0.53 & 0.55 & 0.56 & 0.63 & 0.60 & 0.42 & 0.70 & 0.48 & 0.80 & 0.60 & 0.71 & 0.73 & 0.65±0.12  \\
LogMine & 0.50 & \textbf{1.00} & 0.72 & 0.78 & 0.69 & 0.85 & 0.87 & 0.61 & 0.88 & 0.43 & 0.74 & 0.52 & 0.58 & 0.92 & \underline{0.99} & 0.69 & 0.74±0.18  \\
Logram & 0.85 & 0.70 & 0.74 & \underline{0.98} & 0.96 & 0.97 & \underline{0.97} & 0.46 & 0.67 & 0.85 & 0.55 & 0.95 & 0.90 & 0.76 & 0.96 & 0.96 & 0.83±0.16  \\
LogSig & 0.54 & \underline{0.73} & 0.23 & 0.38 & 0.09 & 0.51 & 0.63 & 0.11 & 0.52 & 0.44 & 0.84 & 0.49 & 0.54 & 0.76 & 0.68 & 0.78 & 0.52±0.23  \\
MoLFI & 0.63 & \textbf{1.00} & 0.95 & 0.51 & 0.46 & \textbf{1.00} & 0.72 & 0.28 & 0.64 & 0.54 & 0.21 & 0.01 & 0.42 & 0.66 & 0.41 & 0.84 & 0.58±0.28  \\
SHISO & 0.58 & \textbf{1.00} & 0.71 & 0.33 & 0.40 & \textbf{1.00} & 0.87 & 0.67 & 0.59 & 0.62 & 0.72 & 0.52 & 0.91 & 0.58 & 0.70 & 0.66 & 0.68±0.19  \\
SLCT & 0.88 & \underline{0.73} & 0.57 & 0.84 & 0.33 & 0.55 & 0.42 & 0.30 & 0.56 & 0.52 & 0.87 & 0.52 & 0.69 & 0.88 & 0.70 & 0.73 & 0.63±0.19  \\
Spell & 0.92 & \textbf{1.00} & 0.79 & \textbf{1.00} & 0.65 & 0.78 & 0.64 & 0.61 & 0.76 & 0.55 & 0.76 & 0.53 & 0.91 & 0.84 & \underline{0.99} & 0.96 & 0.79±0.16  \\
UniParser & \textbf{0.97} & \textbf{1.00} & \textbf{1.00} & \textbf{1.00} & \underline{0.97} & \textbf{1.00} & \textbf{1.00} & 0.88 & \textbf{1.00} & \textbf{1.00} & \textbf{1.00} & 0.98 & \textbf{1.00} & \textbf{0.99} & \textbf{1.00} & \textbf{1.00} & \textbf{0.99±0.03}  \\
LogPPT & 0.89 & \textbf{1.00} & 0.95 & \textbf{1.00} & 0.94 & \underline{0.99} & \textbf{1.00} & \underline{0.93} & 0.78 & 0.63 & \underline{0.99} & \textbf{1.00} & \textbf{1.00} & 0.68 & \underline{0.99} & \underline{0.99} & 0.92±0.12  \\
LILAC & 0.93 & \textbf{1.00} & \underline{0.98} & \textbf{1.00} & \underline{0.97} & \underline{0.99} & \textbf{1.00} & 0.75 & 0.82 & 0.56 & \textbf{1.00} & \textbf{1.00} & \textbf{1.00} & \underline{0.98} & \underline{0.99} & \underline{0.99} & 0.94±0.12  \\
\short{} & \underline{0.94} & \textbf{1.00} & 0.95 & \underline{0.98} & \textbf{1.00} & \textbf{1.00} & 0.96 & \textbf{0.98} & \underline{0.90} & \underline{0.99} & \textbf{1.00} & \underline{0.99} & \textbf{1.00} & 0.96 & \textbf{1.00} & 0.97 & \underline{0.98±0.03}  \\

\bottomrule
\end{tabular}
\end{table*}

\begin{table*}[htbp]
\captionsetup{skip=1pt}
\centering
\setlength{\tabcolsep}{2pt} 
\footnotesize 
\caption{Group Accuracy Comparison on LogHub-2.0. The highest group accuracy for each dataset is highlighted in bold, and the second highest is underlined. Missing data indicates the corresponding method failing to finish.}
\label{tbl:group-accuracy-full}
\begin{tabular}{lccccccccccccccc}
\toprule
Method & Apache & BGL & HDFS & HPC & Hadoop & HealthApp & Linux & Mac & OpenSSH & OpenStack & Proxifier & Spark & Thunderbird & Zookeeper & Average \\
\midrule
AEL & \textbf{1.00} & \textbf{0.92} & \textbf{1.00} & 0.75 & 0.82 & 0.73 & \underline{0.92} & 0.80 & \underline{0.71} & 0.74 & 0.97 & \diagbox{}{} & 0.79 & \textbf{1.00} & 0.86±0.11  \\
Drain & \textbf{1.00} & \textbf{0.92} & \textbf{1.00} & 0.79 & \textbf{0.92} & 0.86 & 0.69 & 0.76 & \underline{0.71} & 0.75 & 0.69 & 0.89 & \textbf{0.83} & \underline{0.99} & 0.84±0.11  \\
IPLoM & \underline{0.99} & 0.90 & \underline{0.96} & 0.79 & \textbf{0.92} & \underline{0.98} & 0.81 & 0.64 & 0.41 & 0.38 & 0.80 & 0.72 & 0.72 & \underline{0.99} & 0.79±0.20  \\
LenMa & \underline{0.99} & \diagbox{}{} & \textbf{1.00} & 0.79 & 0.80 & \diagbox{}{} & 0.81 & 0.70 & \textbf{0.75} & 0.85 & 0.50 & \diagbox{}{} & \diagbox{}{} & 0.86 & 0.81±0.14  \\
LFA & 0.81 & 0.73 & 0.75 & 0.73 & 0.83 & 0.80 & 0.23 & 0.59 & 0.16 & 0.67 & 0.35 & 0.60 & 0.38 & 0.84 & 0.61±0.23  \\
LogCluster & 0.55 & 0.76 & 0.56 & 0.73 & 0.48 & 0.73 & 0.60 & 0.46 & 0.22 & 0.69 & 0.66 & 0.41 & 0.42 & 0.74 & 0.57±0.16  \\
LogMine & \textbf{1.00} & 0.64 & \diagbox{}{} & \diagbox{}{} & 0.83 & \diagbox{}{} & 0.74 & 0.85 & \diagbox{}{} & \diagbox{}{} & 0.50 & \diagbox{}{} & \diagbox{}{} & 0.70 & 0.75±0.16  \\
Logram & 0.30 & \diagbox{}{} & \diagbox{}{} & 0.64 & 0.19 & 0.23 & 0.13 & 0.36 & 0.22 & 0.53 & 0.03 & \diagbox{}{} & \diagbox{}{} & 0.75 & 0.34±0.23  \\
LogSig & 0.11 & \diagbox{}{} & 0.37 & 0.00 & 0.00 & 0.00 & 0.14 & 0.00 & 0.47 & 0.38 & 0.49 & \diagbox{}{} & \diagbox{}{} & 0.00 & 0.18±0.21  \\
MoLFI & 0.59 & \diagbox{}{} & \textbf{1.00} & 0.66 & 0.65 & 0.55 & 0.19 & 0.53 & 0.45 & 0.27 & 0.00 & \diagbox{}{} & \diagbox{}{} & 0.83 & 0.52±0.29  \\
SHISO & 0.57 & 0.60 & \textbf{1.00} & 0.08 & 0.72 & 0.08 & 0.07 & 0.61 & 0.40 & 0.81 & 0.69 & \diagbox{}{} & \diagbox{}{} & 0.82 & 0.54±0.31  \\
SLCT & 0.42 & \diagbox{}{} & 0.41 & 0.64 & 0.23 & 0.11 & 0.08 & 0.42 & 0.28 & \textbf{1.00} & 0.02 & \diagbox{}{} & \diagbox{}{} & 0.75 & 0.40±0.30  \\
Spell & \textbf{1.00} & 0.68 & \underline{0.96} & \diagbox{}{} & 0.45 & 0.65 & 0.62 & 0.76 & 0.62 & 0.78 & 0.52 & \diagbox{}{} & \diagbox{}{} & \underline{0.99} & 0.73±0.19  \\
UniParser & 0.29 & 0.55 & \textbf{1.00} & 0.79 & 0.72 & 0.45 & 0.26 & \underline{0.89} & 0.50 & \textbf{1.00} & 0.51 & 0.85 & 0.44 & \textbf{1.00} & 0.66±0.26  \\
LogPPT & 0.79 & 0.31 & 0.69 & 0.78 & 0.53 & 0.84 & 0.20 & 0.54 & 0.28 & 0.53 & 0.51 & 0.45 & 0.42 & 0.97 & 0.56±0.23  \\
LILAC & \textbf{1.00} & 0.89 & \textbf{1.00} & \textbf{0.87} & \underline{0.87} & \textbf{1.00} & \textbf{0.97} & \textbf{0.90} & 0.69 & \textbf{1.00} & \textbf{1.00} & \textbf{1.00} & \underline{0.81} & \textbf{1.00} & \textbf{0.93±0.10}  \\
\short{} & \underline{0.99} & \underline{0.91} & \textbf{1.00} & \underline{0.80} & \textbf{0.92} & 0.96 & 0.81 & 0.81 & 0.63 & \underline{0.99} & \underline{0.98} & \underline{0.97} & 0.78 & 0.97 & \underline{0.90±0.11}  \\
\bottomrule
\end{tabular}
\end{table*}

We rigorously evaluate the effectiveness of our method, \short{}, by comparing it with state-of-the-art log parsing approaches across two diverse datasets: 16 small labeled datasets from LogHub and 14 large-scale datasets from LogHub-2.0. 
The results, presented in \cref{tbl:group-accuracy-2k} and \cref{tbl:group-accuracy-full}, demonstrate \short{}'s superior performance across a wide spectrum of log types and volumes.

On the LogHub dataset, \short{} achieved an impressive average grouping accuracy of 0.98, surpassing most existing methods. 
It consistently ranked among the top performers across individual datasets, showcasing its versatility in handling various log patterns effectively. 
\short{}'s performance was particularly notable on complex datasets (e.g., Linux and Mac), underscoring its robustness and adaptability to diverse log structures.

The advantages of \short{} became even more pronounced when evaluated on the large-scale LogHub-2.0 datasets. 
With an average grouping accuracy of 0.90, \short{} significantly outperformed most baselines, many of which experienced substantial performance degradation when scaling to larger log volumes. 
Notably, \short{} maintained high accuracy across different log types, from system logs (e.g., HDFS, Spark) to application logs (e.g., Thunderbird, HealthApp), demonstrating its scalability and consistency in handling massive and diverse log data. 
While LILAC showed slightly higher accuracy, its poor efficiency (as detailed in \cref{sec:effectiveness-comparison}) limits its practical applicability in real-world scenarios, especially for large-scale, real-time log parsing tasks.

It's worth noting that \short{}'s performance remained consistently high across datasets of varying sizes and complexities, despite not being specifically optimized for any particular log type or domain. 
This domain-agnostic effectiveness demonstrates \short{}'s inherent ability to handle diverse and unpredictable log patterns, a crucial attribute for a cloud-based log parsing service that must process logs from various applications and systems.
Moreover, \short{} successfully completed parsing tasks on all datasets, unlike some competing methods that failed to finish on certain large-scale logs, further highlighting its robustness and reliability.
In some datasets, such as Mac, our method shows slightly lower accuracy. This is primarily because our approach is syntax-based, making it difficult to capture patterns in logs that require a holistic semantic understanding of structured text.

Overall, these comprehensive results solidify the effectiveness of our log parsing method.

\subsection{Efficiency Comparison}
\label{sec:efficiency-comparison}

\begin{figure*}[htbp]
\captionsetup{skip=1pt}
    \centering
    \includegraphics[width=\linewidth]{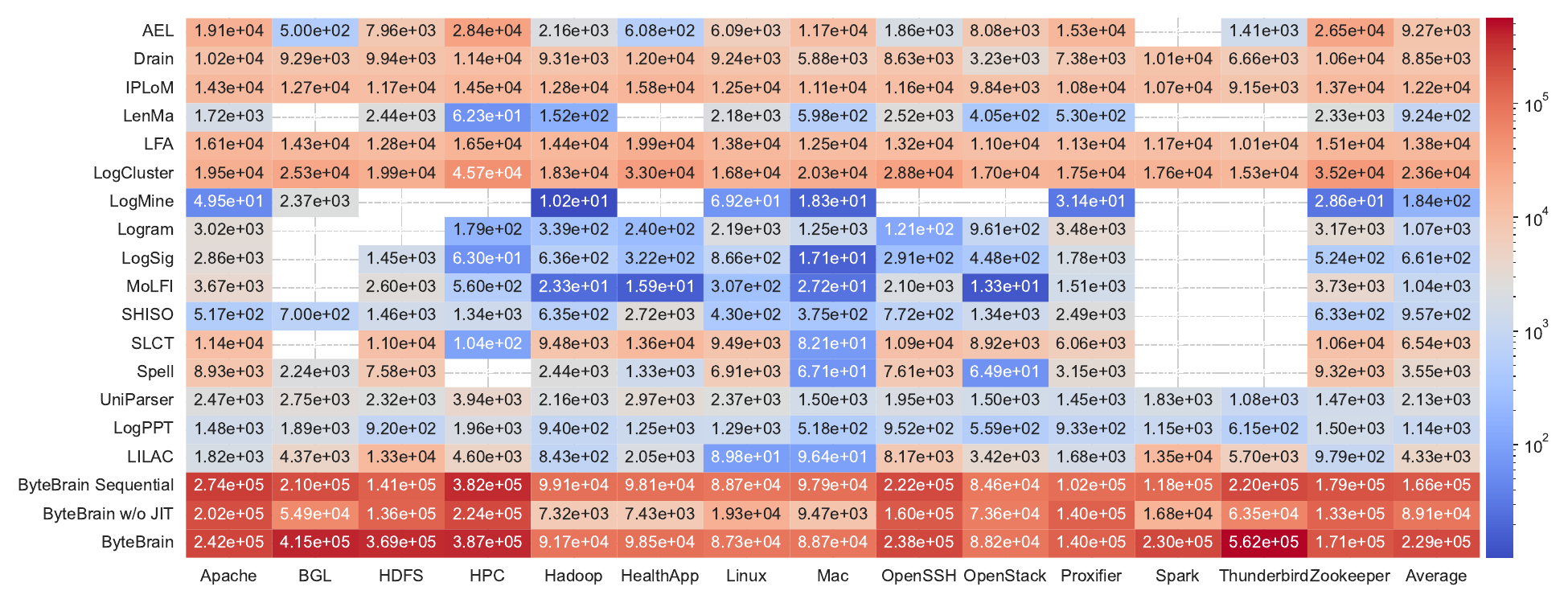}
    \caption{Throughput (\#logs/second) comparison on LogHub-2.0. Missing data indicates failing to finish.}
    \Description{A scatter plot comparing group accuracy (y-axis, 0.2-0.8) versus training throughput (x-axis, 10^3-10^5 logs/second, logarithmic scale) across 16 different log parsing methods. \short{} achieves the highest throughput while maintaining comparable accuracy to state-of-the-art methods, significantly outperforming both traditional syntax-based approaches and modern deep learning-based methods.}
    \label{fig:throughput-comparison}
\end{figure*}
We conducted the throughput experiments on a server equipped with an Intel(R) Xeon(R) Platinum 8336C CPU @ 2.30GHz and 128GB RAM. 
Our algorithm, including both training and matching, is implemented in Python, leveraging Just-In-Time (JIT) compilation for code acceleration and multi-threading parallel processing.

As shown in \cref{fig:throughput-comparison}, our method significantly outperforms all baseline methods in terms of throughput across the majority of datasets. 
On average, \short{} achieves a remarkable throughput of 229 thousand logs per second, which is 1-3 orders of magnitude higher than most existing methods and 840.68\% faster than the fastest baseline (LogCluster).
This exceptional performance is particularly evident in large-scale datasets such as Thunderbird, where our method processes 519 thousand logs per second, far surpassing the next best performer.

Among the baseline methods, traditional approaches like AEL, Drain, and IPLoM show relatively consistent performance across datasets, with average throughputs ranging from 8,850 to 12,200 logs per second. However, more sophisticated methods such as UniParser and LogPPT, while potentially offering higher accuracy, suffer from significantly lower throughput, processing only 2,130 and 1,140 logs per second on average, respectively.
Notably, some methods like LenMa, LogMine, and Logram failed to complete parsing on several datasets, indicating limitations in their efficiency or ability to handle diverse log formats.

To ensure a fair comparison, we also evaluated our method's throughput when using a single core (\short{}  Sequential) and when Just-In-Time (JIT) compilation is disabled (\short{} w/o JIT, which also disables parallelization as multi-threading is not available without JIT). 
Even with a single core, our method maintains an impressive average throughput of 166,000 logs per second, significantly outperforming all baseline methods. 
The modest performance gain of \short{} over \short{} Sequential is expected and consistent with our findings that parallelism offers limited improvement on smaller datasets, as demonstrated in our parallelism scalability analysis (see \cref{fig:throughput-vs-parallelism}).
Furthermore, when JIT compilation is disabled, although the throughput decreases to 89,100 logs per second, it still surpasses baseline methods by at least an order of magnitude. 
These results demonstrate that the superior performance of our method stems from its algorithmic efficiency rather than merely from optimized implementation.

\begin{figure}[t]
\captionsetup{skip=1pt}
    \centering
    \includegraphics[width=\linewidth]{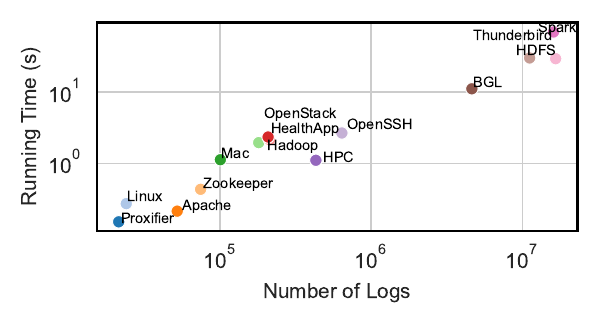}
    \caption{Our running time scales linearly with log size}
    \Description{A log-log scatter plot showing the relationship between running time (y-axis, 0.1-100 seconds) and number of logs (x-axis, 10^4-10^7) when running ByteBrain-LogParser on different LogHub-2.0 datasets. The plot reveals a near-linear relationship between processing time and log volume across 13 datasets (including Proxifier, Linux, Apache, etc.), demonstrating the algorithm's consistent scalability regardless of dataset size or type. Most datasets form a clear trend line, with running time increasing proportionally with log volume.}
    \label{fig:running-time-vs-logs}
\end{figure}

\Cref{fig:running-time-vs-logs} provides further insight into \short{}'s efficiency by plotting the running time against the number of logs for different datasets. 
The graph shows a near-linear relationship between processing time and log volume, with most datasets clustered along a similar trajectory. 
This linear scaling demonstrates our method is able to efficiently handle increasing log volumes, which is a crucial feature for cloud-based log parsing services.

In summary, these efficiency comparisons highlight the superior performance of our method in processing large-scale log data. 
The high throughput and linear complexity make it particularly well-suited for cloud-based log parsing services, where handling massive volumes of diverse log data efficiently is crucial.

\subsection{Ablation Study}
\label{sec:ablation-study}

We compare \short{} with the following variants with respect to either accuracy or efficiency to validate the effectiveness of our proposed techniques:
\begin{itemize}[leftmargin=1em]
\item \textit{w/ naive match}: Use the templates assigned during clustering for training logs instead of the matching method in \cref{sec:online-matching}.
\item \textit{w/o position importance}: For positional similarity distance, set $w_i=1$ in \cref{eq:positional-similarity-distance}.
\item \textit{w/o variable in saturation}: For saturation, set $s(C) = f_c$ in \cref{eq:saturation}.
\item \textit{w/o confidence factor}: For saturation, set $s(C) = f_v \cdot f_c$ in \cref{eq:saturation}.
\item \textit{random centroid selection}: Randomly select initial centroids for new clusters instead of the K-Means++ strategy.
\item \textit{w/o ensure saturation increase}: Split each node into two clusters, even if their saturation scores do not increase.
\item \textit{w/o balanced group} and \textit{w/o early stopping}
\item \textit{w/o deduplication \& related techs}: Skip deduplication and dependent optimizations like balanced group and early stopping.
\end{itemize}

\begin{figure}[t]
\captionsetup{skip=1pt}
\centering
\includegraphics[width=\linewidth]{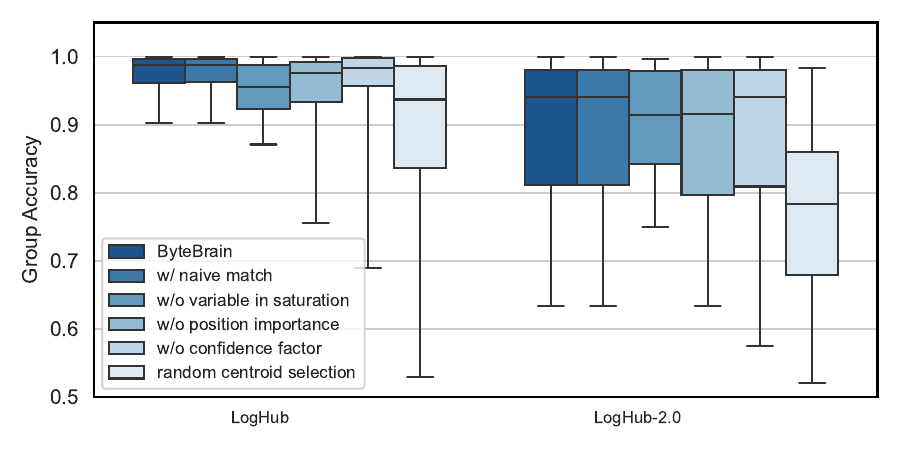}
\caption{Our online match method maintains performance, while other techniques improve accuracy.}
\label{fig:ablation-effectiveness}
\Description{This figure shows the group accuracy comparison for ByteBrain and its ablation variants on the LogHub and LogHub-2.0 datasets. The box plots represent the distribution of group accuracy for each method. ByteBrain achieves the highest and most consistent accuracy, while methods with removed components (e.g., "w/o variable in saturation" or "random centroid selection") show reduced accuracy and greater variability, especially on the more complex LogHub-2.0 dataset.}
\end{figure}

\subsubsection{Text-based matching does not compromise accuracy}
As shown in \cref{fig:ablation-effectiveness}, directly using the templates assigned to training logs during clustering produces almost identical group accuracy compared to assigning templates by matching each log with the template texts after the clustering tree is built. 
Therefore, the online matching method proposed in \cref{sec:online-matching} significantly reduces storage overhead with virtually no impact on matching performance.

\subsubsection{Accuracy improvement brought by the proposed techniques}
As shown in \cref{fig:ablation-effectiveness}, removing variable positions from saturation calculation reduces accuracy, confirming its contribution. Interestingly, on LogHub-2.0, \textit{w/o variable saturation} achieves higher minimum accuracy, suggesting our heuristic position estimation may occasionally misclassify positions on challenging datasets. Nevertheless, variable saturation consistently improves overall performance across all datasets.

Removing position importance from distance calculations decreases accuracy, confirming its value in capturing structural significance of log positions. This effect is more pronounced on the larger, more complex LogHub-2.0 datasets, underscoring these techniques' importance when handling diverse, voluminous log data.

Random centroid selection causes the most severe accuracy reduction, highlighting intelligent selection's critical role in achieving high parsing accuracy, especially for large-scale, diverse log datasets.
The confidence factor removal from saturation calculation shows relatively minor impact across datasets, indicating its less significant contribution compared to other techniques.

This ablation study confirms each technique contributes to \short{}'s accuracy, with their importance amplified by increasing log data scale and complexity. Together, these techniques enable \short{} to maintain robust performance across diverse log types and volumes.

\subsubsection{Efficiency improvement brought by the proposed techniques}
\begin{figure}[t]
\captionsetup{skip=1pt}
	\includegraphics[width=\linewidth]{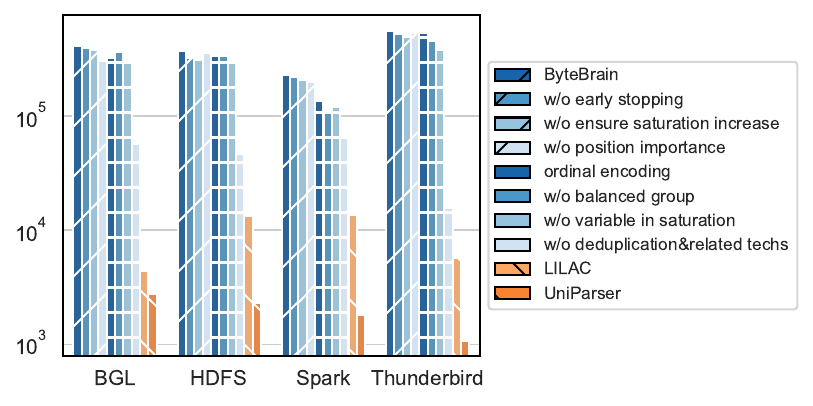}
	\caption{Our method boosts throughput and remains much faster than baselines without individual techniques.}
	\label{fig:throughput-ablation}
    \Description{A bar chart with logarithmic y-axis (10^3 to 10^5) showing throughput comparisons across four large-scale datasets (BGL, HDFS, Spark, and Thunderbird). The chart compares ByteBrain against its variants with different features disabled and two baseline methods (LILAC and UniParser). Each dataset has 10 bars, with ByteBrain showing the highest throughput consistently around 10^5 logs/second. The variants show progressively decreasing performance as key features are disabled, with "w/o deduplication&related techs" showing the most significant drop. Baseline methods LILAC and UniParser show the lowest throughput at around 10^3-10^4 logs/second. This visualization demonstrates the cumulative impact of ByteBrain's optimization techniques on processing speed.}
\end{figure}

As shown in \cref{fig:throughput-ablation}, our proposed techniques deliver significant efficiency improvements across the four large-scale datasets (over 500MB).

The most striking impact comes from deduplication and its related techniques. 
Without these optimizations, the throughput drops dramatically, particularly evident in the Thunderbird dataset where performance decreases by about two orders of magnitude. 
This underscores the critical role of deduplication, balanced grouping, and early stopping in handling large-scale log data efficiently.
Nevertheless, when compared to the best performing baselines like LILAC and UniParser, the throughput of each variant is consistently higher by one or two orders of magnitude across all datasets. 

Variable saturation scoring is the second most important improvement, enabling faster convergence by reducing unnecessary splits on variable positions during training.   
Balanced grouping ranks third, preventing unbalanced clusters where a single node dominates, ensuring efficient processing across the clustering tree.  
Hash encoding also provides a notable speedup by enabling parallel token processing, unlike ordinal encoding, which requires sequential mapping.  
Other proposed techniques, such as position importance and ensuring saturation increase, also demonstrate noticeable improvements in throughput across different datasets, though to a lesser extent.

Importantly, while some techniques yield limited improvements on specific datasets, they consistently contribute positively across scenarios without any performance degradation.  
By combining all these techniques, our method achieves remarkable performance across all datasets.

\subsubsection{Hash encoding reduces space consumption}
\label{sec:hash-encoding-reduce-space-consumption}
\begin{figure}[t]
\captionsetup{skip=1pt}
    \centering
    \includegraphics[width=\linewidth]{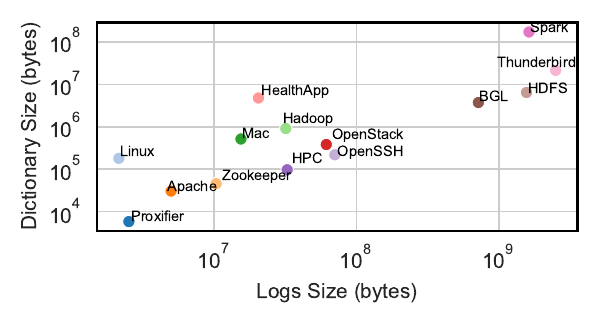}
    \caption{Dictionary size with ordinal encoding, demonstrating storage savings by hash encoding}
    \Description{A log-log scatter plot showing the relationship between dictionary size required by ordinal encoding (y-axis, 10^4-10^8 bytes) and total log size (x-axis, 10^7-10^9 bytes) across LogHub-2.0 datasets. Each dataset is represented by a colored dot. The plot demonstrates that larger log datasets like Spark and Thunderbird (around 10^9 bytes) require substantially larger dictionaries (around 10^8 bytes), while smaller datasets like Proxifier need much smaller dictionaries (around 10^4 bytes). This visualization highlights the significant storage overhead of ordinal encoding that ByteBrain's hash encoding approach eliminates.}
    \label{fig:dictionary-size}
\end{figure}
We study the sizes of the dictionary files (mapping of tokens to encodings) generated by ordinal encoding on LogHub-2.0 datasets, which represent the space savings we achieve by using hash encoding.
As shown in \cref{fig:dictionary-size}, as log size increases, the dictionary size required for ordinal encoding grows significantly, reaching hundreds of megabytes for large datasets like Thunderbird and Spark. 
Our hash encoding method eliminates the need for storing these large dictionaries entirely. 
This approach not only reduces storage requirements but also improves parsing efficiency by minimizing data transfer overhead.
Consequently, it helps users minimize their operational costs in cloud environments, where storage incurs ongoing expenses.
Moreover, the space savings become increasingly significant as the scale of log data grows, making our method particularly suitable for large-scale, cloud-based log parsing services.

\subsection{Parameter Sensitivity}
\label{sec:parameter-sensitivity}
\subsubsection{Saturation}
\begin{figure*}[htbp]
\captionsetup{skip=1pt}
\centering
\includegraphics[width=\linewidth]{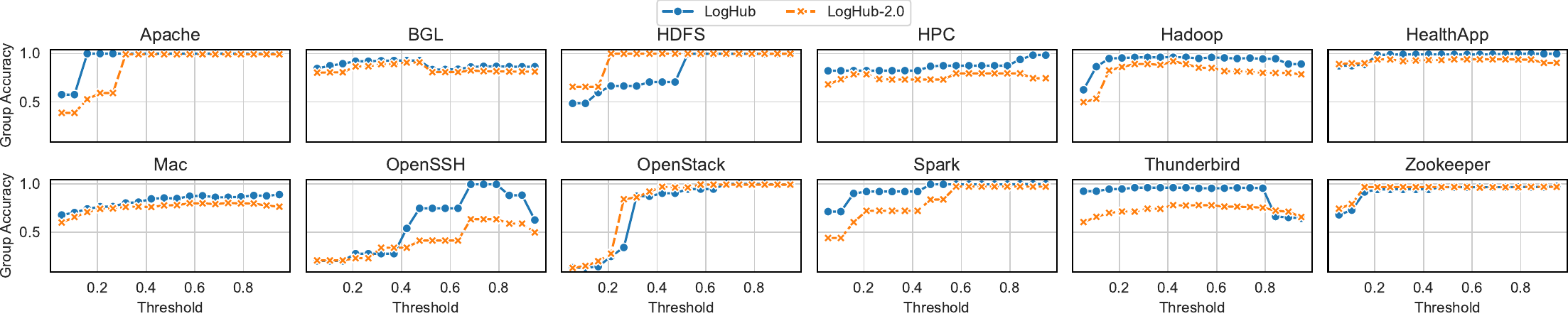}
\caption{Group accuracy with different saturation threshold on LogHub and LogHub-2.0 (selected datasets due to space limit)}
\Description{A series of line graphs showing group accuracy (y-axis, 0-1.0) versus saturation threshold (x-axis, 0-1.0) for twelve datasets from LogHub and LogHub-2.0. Each subplot represents a different dataset (Apache, BGL, HDFS, HPC, Hadoop, HealthApp, Mac, OpenSSH, OpenStack, Spark, Thunderbird, and Zookeeper), with two lines: blue circles for LogHub and orange crosses for LogHub-2.0. The plots demonstrate that ByteBrain-LogParser maintains relatively stable accuracy across different threshold values for most datasets, with some variation between the smaller LogHub and larger LogHub-2.0 versions. This visualization illustrates the robustness of the parsing algorithm across different saturation thresholds and dataset sizes.}
\label{fig:group-accuracy-threshold}
\end{figure*}

As shown in \cref{fig:group-accuracy-threshold}, the group accuracy of our method remains relatively stable across a wide range of saturation thresholds, indicating that our method is not overly sensitive to this parameter. 
This robustness ensures consistent parsing results even when the threshold is not precisely tuned.

\begin{table}[t]
\captionsetup{skip=1pt}
\centering
\setlength{\tabcolsep}{1.5pt} 
\footnotesize 
\caption{Templates obtained by varying saturation thresholds showing adaptability}
\label{tbl:template-examples}
\begin{tabular}{crrrrrrrrrrrrrrr}
\toprule
Saturation            & \multicolumn{15}{c}{Template}                                                                 \\ \midrule
0.05                  & *       & lock & * & *     & * & tag & * & name & *           & ws & *    & uid & * & pid & * \\ \midrule
\multirow{2}{*}{0.78} & release & lock & * & flg   & * & tag & * & name & *           & ws & *    & uid & * & pid & * \\
                      & acquire & lock & * & flags & * & tag & * & name & *           & ws & *    & uid & * & pid & * \\ \midrule
\multirow{4}{*}{0.9}  & release & lock & * & flg   & * & tag & * & name & android     & ws & *    & uid & * & pid & * \\
                      & release & lock & * & flg   & * & tag & * & name & *           & ws & null & uid & * & pid & * \\
                      & acquire & lock & * & flags & * & tag & * & name & android     & ws & *    & uid & * & pid & * \\
                      & acquire & lock & * & flags & * & tag & * & name & *           & ws & null & uid & * & pid & * \\ \midrule
\multirow{8}{*}{0.95} & release & lock & * & flg   & * & tag & * & name & android     & ws & *    & uid & * & pid & * \\
                      & release & lock & * & flg   & * & tag & * & name & *           & ws & null & uid & * & pid & * \\
                      & release & lock & * & flg   & * & tag & * & name & audioserver & ws & null & uid & * & pid & * \\
                      & release & lock & * & flg   & * & tag & * & name & android     & ws & null & uid & * & pid & * \\
                      & acquire & lock & * & flags & * & tag & * & name & android     & ws & null & uid & * & pid & * \\
                      & acquire & lock & * & flags & * & tag & * & name & android     & ws & *    & uid & * & pid & * \\
                      & acquire & lock & * & flags & * & tag & * & name & audioserver & ws & null & uid & * & pid & * \\
                      & acquire & lock & * & flags & * & tag & * & name & *           & ws & null & uid & * & pid & * \\
                      \bottomrule
\end{tabular}
\end{table}

While our method maintains relatively stable performance across different saturation thresholds, the threshold still effectively controls template precision when varied across a wider range.
This controllable behavior is desirable, as it allows users to adjust parsing precision according to their needs while maintaining robustness against small parameter perturbations.

To illustrate how the saturation threshold affects template generation, \cref{tbl:template-examples} shows templates produced at different threshold values for Android logs in LogHub.
At a low threshold (0.05), the template is highly generalized with most fields marked as variables (*).
As the threshold increases to 0.78, more structural elements like "release" and "acquire" are preserved.
At higher thresholds (0.9 and 0.95), the templates become increasingly specific, distinguishing between similar terms like "flg" and "flags" and preserving system-specific values like "android" and "audioserver".
This progression demonstrates how users can effectively control template granularity to support different analysis scenarios, from high-level pattern recognition to detailed debugging.

\subsubsection{Parallelism}
\begin{figure*}[htbp]
\captionsetup{skip=1pt}
\centering
\includegraphics[width=\linewidth]{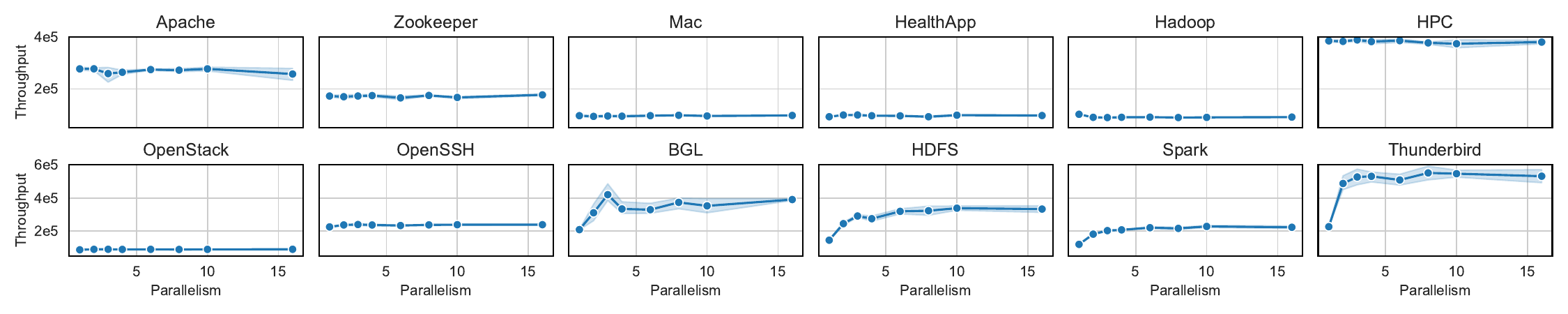}
\caption{Throughput vs/ parallelism on LogHub-2.0 (selected datasets due to space limit). The datasets are sorted by size.}
\Description{A collection of twelve line plots showing the relationship between throughput (y-axis) and degree of parallelism (x-axis, 1-15 threads) for different LogHub-2.0 datasets. The datasets (Apache, Zookeeper, Mac, HealthApp, Hadoop, HPC, OpenStack, OpenSSH, BGL, HDFS, Spark, and Thunderbird) are arranged by size. Each plot shows throughput values with blue dots connected by lines, with a light blue confidence interval. Most datasets show an initial increase in throughput with increased parallelism before plateauing, though the effect varies by dataset size. Smaller datasets show modest gains, while larger datasets like Thunderbird demonstrate more substantial throughput improvements with increased parallelism.}
\label{fig:throughput-vs-parallelism}
\end{figure*}
We investigated the impact of parallelism on throughput.
In \cref{fig:throughput-vs-parallelism}, as the degree of parallelism increases, we see a general trend of improved throughput across large-scale datasets, while smaller datasets like Linux and Proxifier show relatively modest gains with increased parallelism. 
This suggests that \short{}'s parallel processing capabilities are particularly beneficial for handling large-scale log data. 
Interestingly, we observe that the throughput improvement tends to plateau as parallelism increases beyond a certain point, especially for smaller datasets. 
This indicates that there's an optimal level of parallelism for each dataset size, beyond which additional parallel processing may not yield significant performance gains. 

\section{Industrial Evaluation}
\label{sec:industrial-evaluation}
\short{} has been successfully deployed as part of Volcano Engine's Torch Log Service (TLS) and is now available to invited users.
In production, we use a Go-based service to schedule training tasks, which are executed on separated Pods and utilize the same Python implementation as described in \cref{sec:experiment}.
For online matching, which must integrate with conventional text indexing systems, we reimplemented the matching module in C++ and Rust and embedded it directly into the log indexing pipeline.
This strategic integration eliminates cross-server data transfer (between C++ indexing code and our log parsing module), substantially reducing I/O overhead and log parsing latency.

The system enables users to organize queried logs by their corresponding templates, providing a structured and intuitive view of complex log data.
A distinctive feature is the real-time precision adjustment capability through an interactive slider in the web interface, allowing users to dynamically control template granularity based on specific analytical requirements.
This interactive capability helps users identify patterns and anomalies more effectively by allowing them to switch between different levels of abstraction on demand.
Based on the parsed log templates, users can save selected templates to a template library, which can then be used to configure alerts (e.g., sudden changes in template count or the appearance of new templates). 
Users can also compare the templates generated across different time periods to analyze changes in log patterns.

\begin{table}[t]
\captionsetup{skip=3pt}
\centering
\footnotesize
\setlength{\tabcolsep}{2pt}
\caption{Performance evaluation using actual production data from the TLS on Volcano Engine, demonstrating \short{}'s effectiveness in actual cloud environments.}
\label{tbl:production-statistics}
\begin{tabular}{lrrrr}
\hline
Topic Scenario                              & Log Volume & Model Size & Training Time  \\ \hline
Text stream processing & 189 MB/s   & 3 MB       & 0.91s   \\
Webserver access log                  & 57.8 MBs   & 10 MB      & 7.98s   \\
Webserver access log                  & 47.7 MB/s  & 3 MB       & 1.02s   \\
Go HTTP API server                    & 3.51 MB/s  & 7 MB       & 1.65s  \\
Go search server                      & 2.46 MB/s  & 7 MB       & 4.64s   \\ \hline
\end{tabular}
\end{table}

\Cref{tbl:production-statistics} presents performance metrics collected from diverse log topics in production environments, representing characteristic real-world deployment scenarios.
As illustrated, the system processes exceptionally high log volumes, reaching up to 189 MB/s, which is equivalent to millions of logs per second.
Despite this substantial throughput, our algorithm completes model training sessions within seconds, demonstrating remarkable computational efficiency.
The system maintains an end-to-end visible latency of approximately 5-15 seconds per log (encompassing ingestion through completed indexing, enabling user queries), confirming that our log parsing method effectively keeps pace with real-time log generation rates.
It is worth noting that this latency figure includes both log parsing and traditional text indexing operations.
This level of responsiveness satisfies the requirements for most real-time application scenarios.

The results in \Cref{tbl:production-statistics} also highlight the storage efficiency of our method.  
The size of log parsing model for each topic is only a few megabytes, which is significantly smaller than the size of the corresponding raw log text.  
This minimal model size ensures that our approach introduces negligible storage overhead, making it highly cost-efficient for large-scale cloud applications.

Due to privacy constraints preventing direct access to user logs, we cannot determine optimal thresholds for calculating desired template counts and parsing accuracy for each topic.
Nevertheless, we observe that the number of most precise templates (saturation $\geq 0.9$) typically ranges between 1,000 and 10,000 across different deployment scenarios.

The successful integration of \short{} within Volcano Engine demonstrates its practical value in production environments, where it effectively combines high-throughput processing capabilities with flexible, user-centric template management.
This operational validation complements our experimental findings and confirms \short{}'s suitability for enterprise-scale log parsing applications.

\section{Discussion}
\label{sec:discussion}
In this section, we discuss several limitations of our approach and potential improvements that can be made in the future.

First, while our approach achieves excellent accuracy and efficiency, it inherently lacks semantic understanding of log content.
Unlike human operators who naturally group logs based on their meanings, our syntax-based approach can only rely on structural similarities.
Semantic parsers leveraging natural language processing techniques can interpret logs like humans do, intelligently grouping similar logs and separating different ones, though at higher computational costs.
Our choice of a syntax-based approach prioritizes compute-efficiency and cost-efficiency, enabling efficient processing of massive log volumes while maintaining high accuracy.
Looking forward, we may combine semantic-based understanding with syntax-based methods to achieve both efficiency and semantic understanding in future work, creating a hybrid solution that balances the strengths of both approaches.

The second limitation of our approach, similar to other syntax-based methods~\cite{wangSPINEScalableLog2022,heDrainOnlineLog2017}, is the inability to directly parse logs of varying lengths into the same template.
This limitation stems from syntax-based methods relying solely on comparing tokens at identical positions to determine log similarity, without understanding the semantic meaning.
This becomes particularly challenging when logs contain variable-length elements, such as when a print statement outputs a dynamic list, resulting in semantically similar logs being parsed into different templates despite originating from the same log statement.
In this paper, we deliberately choose not to implement dynamic matching solutions (e.g., using longest common subsequence to compare logs of different lengths) to address this challenge.
The reason is that allowing wildcards to match variable numbers of tokens would require a search process during online matching to determine the optimal token spans for each wildcard.
Such an approach would significantly increase the computational complexity of online matching, making it impractical in cloud environments where millions of logs need to be matched per second.
Instead, we propose a simple yet effective optimization at the query result processing layer.
For example, consider three templates generated from the statement \texttt{print(f'users=\{users\}')} where the \texttt{users} list contains one, two, and three elements: \texttt{users *}, \texttt{users * *}, and \texttt{users * * *}.
When processing query results before presentation, we merge consecutive wildcards in each template, resulting in \texttt{users *} for all three cases.
We then group logs with the same merged template together in the response.
This optimization balances user experience and system performance: users see one intuitive template \texttt{users *} that accommodates variable-length lists, while our underlying system maintains efficiency by using the original fixed-length templates during log parsing and matching.

\section{Conclusion}

This paper presents an adaptive and efficient log parsing approach, which is optimized as a cloud service, for processing diverse, massive logs from different tenants. 
Our method offers real-time adjustable parsing precision, incorporates efficiency-enhancing techniques (deduplication, balanced grouping, early termination), and minimizes storage overhead through hash encoding and text-based matching.
Comprehensive evaluations on public datasets demonstrate state-of-the-art efficiency with comparable accuracy, while production deployment validates its practical effectiveness and efficiency.
By integrating out-of-the-box log parsing and intelligent analysis capabilities into our log service, we enhances its overall intelligence and usability.
Future work will extend the framework to handle more complex log patterns, including structural content and dynamic-length variables, enabling broader applications.

\clearpage{}

\bibliographystyle{ACM-Reference-Format}
\bibliography{refs-lzy-zotero-library}  

\end{document}